\documentclass[runningheads]{llncs}

\usepackage{amssymb}
\usepackage{multirow,setspace,verbatim,amsfonts,amsmath,amsbsy,epsfig,url}
\usepackage{gensymb}
\usepackage{booktabs}

\usepackage{cite}
\usepackage{makecell}
\usepackage{epstopdf}

\usepackage{algorithm}
\usepackage[noend]{algpseudocode}
\usepackage{tikz}
\usepackage{mathrsfs} 
\usepackage[algo2e]{algorithm2e}
\usepackage{array}
\usepackage{color}
\usepackage{graphicx}
\usepackage{dsfont}

\usepackage{framed,multirow}
\usepackage{tabularx}
\usepackage{amssymb}
\usepackage{latexsym}
\usepackage{url}
\usepackage{xcolor}
\usepackage{hyperref}

\newcommand{\Sc}{\mathcal{S}}
\newcommand{\Tc}{\mathcal{T}}
\newcommand{\Xc}{\mathcal{X}}
\newcommand{\Yc}{\mathcal{Y}}
\newcommand{\Zc}{\mathcal{Z}}

\usepackage{array,multirow}
\newcolumntype{P}[1]{>{\centering\arraybackslash}p{#1}}
\newcolumntype{M}[1]{>{\centering\arraybackslash}m{#1}}
\newcommand{\beginsupplement}{%
        \setcounter{table}{0}
        \renewcommand{\thetable}{S\arabic{table}}%
        \setcounter{figure}{0}
        \renewcommand{\thefigure}{S\arabic{figure}}%
        \setcounter{section}{0}
        \renewcommand{\thesection}{S\arabic{section}}%
     }
\definecolor{newcolor}{rgb}{.8,.349,.1}

\usepackage{xurl}
\usepackage{orcidlink}

\usepackage{xr-hyper}
\usepackage{hyperref}
\begin{document}
\pagestyle{headings}
\mainmatter
\def\ECCVSubNumber{6338}  

\title{CXR Segmentation by AdaIN-based Domain Adaptation and Knowledge Distillation} 

\titlerunning{ECCV-22 submission ID 6338} 
\authorrunning{ECCV-22 submission ID 6338} 
\author{Anonymous ECCV submission}
\institute{Paper ID 6338}

\titlerunning{CXR Segmentation by AdaIN-based DA and KD}
%
\author{Yujin Oh\orcidlink{0000-0003-4319-8435} \and
Jong Chul Ye\orcidlink{0000-0001-9763-9609} }
\authorrunning{Y. Oh et al.}
%

\institute{Kim Jaechul Graduate School of AI, Korea Advanced Institute of Science and Technology (KAIST), Daejeon, South Korea
\email{jong.ye@kaist.ac.kr}}

\maketitle

\begin{abstract}
As segmentation labels are scarce, extensive researches have been conducted to train segmentation networks with domain adaptation, semi-supervised or self-supervised learning techniques to utilize abundant unlabeled dataset. However, these approaches appear different from each other, so it is not clear how these approaches can be combined for better performance. Inspired by recent multi-domain image translation approaches, here we propose a novel segmentation framework using adaptive instance normalization (AdaIN), so that a single generator is trained to perform both domain adaptation and semi-supervised segmentation tasks via knowledge distillation by simply changing task-specific AdaIN codes. Specifically, our framework is designed to deal with difficult situations in chest X-ray radiograph (CXR) segmentation, where labels are only available for normal data, but the trained model should be applied to both normal and abnormal data. The proposed network demonstrates great generalizability under domain shift and achieves the state-of-the- art performance for abnormal CXR segmentation.
\end{abstract}

\section{Introduction}

High-accuracy image segmentation often serves as the first step in various medical image analysis tasks \cite{de2018clinically, ouyang2020dual}.
Recently, deep learning (DL) approaches have  become the state-of-the-art (SOTA) techniques for medical image segmentation tasks thanks to their superior performance compared to the classical methods \cite{de2018clinically}. 

The performance of DL-based segmentation algorithm usually depends on large amount of labels, but segmentation masks are scarce due to expensive and time-consuming annotation procedures. 
Another difficulty in DL-based segmentation is the so-called domain shift, i.e., a segmentation network trained with data in a specific domain  often undergoes drastic performance degradation when applied to unseen test domains.  

{For example, in the field of chest X-ray radiograph (CXR) analysis \cite{ccalli2021deep}, segmentation networks trained with normal CXR data often produce under-segmentation when applied to abnormal CXRs with severe infectious diseases such as viral or bacterial pneumonia \cite{y2020deep, signoroni2021bs}. 
The missed regions from under-segmentation mostly contain crucial features, such as pulmonary consolidations or ground-glass opacity, for classifying the infectious diseases.}
Thus, highly-accurate lung segmentation results without under-segmentation are required to guarantee that DL-based classification algorithms fully learn crucial lung features, while alleviating irrelevant factors outside the lung to prevent shortcut learning \cite{degrave2021ai}.

To solve the label scarcity and domain shift problems, there have been extensive researches to train segmentation networks in a semi-supervised manner using limited training labels, or  in a self-supervised or unsupervised manner even without labeled dataset \cite{bai2017semi, tang2019XLSor, tarvainen2017mean, perone2019unsupervised, xue2020dual, orbes2019multi, li2020transformation, chen2020unsupervised}. 
However, these approaches appear different from each other, and there exist no consensus in regard to how these different approaches can be synergistically combined.
Inspired by success of StarGANv2 for transferring style between various domains  \cite{choi2020stargan}, here we propose a style transfer-based knowledge distillation framework that can synergistically combine supervised segmentation, domain adaptation and self-supervised knowledge distillation tasks to improve unsupervised segmentation performance. 
Specifically, our framework is designed to deal with difficult but often encountered situations where segmentation masks are only available for normal data, but the trained method should be applied to both normal and abnormal images.

\begin{figure*}[tb!]
\centering
\includegraphics[width=12cm]{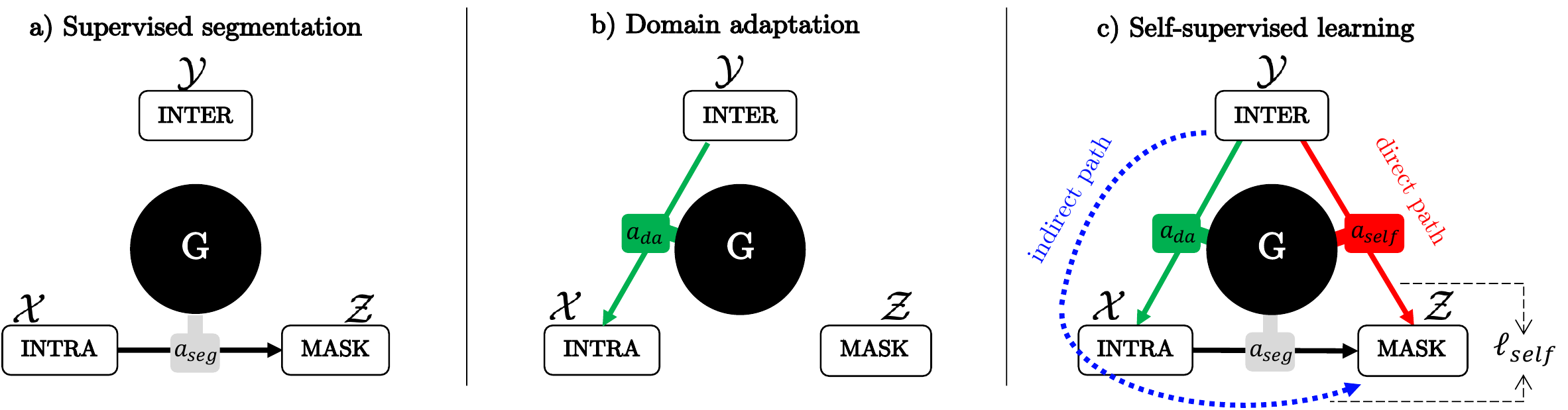}
\caption{Overview of the proposed unified framework. A single generator G transfers domains by simply changing task-specific AdaIN codes ${a_{seg}}$, ${a_{da}}$ and ${a_{self}}$ for each supervised segmentation, domain adaptation and self-supervised learning task, respectively.}
\label{fig:scenario}
\end{figure*}

The key idea is that a single generator trained along with the adaptive instance normalization (AdaIN) \cite{huang2017arbitrary} can perform supervised segmentation as well as domain adaptation between normal and abnormal domains by simply changing the AdaIN codes, as shown in Fig. \ref{fig:scenario}(a) and (b), respectively.
The network is also trained in a self-supervised manner using another AdaIN code in order to force direct segmentation results (illustrated as a red arrow in Fig. \ref{fig:scenario}(c)) to be matched to indirect segmentation results through domain adaptation and subsequent segmentation (illustrated as a blue arrow in Fig. \ref{fig:scenario}(c)). 
Since a single generator is used for all these tasks, the network can synergistically learn common features from different dataset through knowledge distillation. 
  
To validate the concept of the proposed framework, we train our network using labeled normal CXR dataset and unlabeled pneumonia CXR dataset, and test the model performance on unseen dataset composed of COVID-19 pneumonia CXRs \cite{signoroni2021bs, vaya2020bimcv}.
{We further evaluate the network performance on domain-shifted CXR datasets for both normal and abnormal cases, and compared the results with other SOTA techniques.
Experimental results confirm that our method has great promise in providing robust segmentation results for both in-domain and out-of-domain dataset.}
We release our source code \footnote{\url{https://github.com/yjoh12/CXR-Segmentation-by-AdaIN-based-Domain-Adaptation-and-Knowledge-Distillation.git}} for being utilized in various CXR analysis tasks presented in Supplementary Section \ref{sp:applications}.

\section{Related Works}
\label{sec:related}

To make this paper self-contained, here we review several existing works that are necessary to understand our method.

\subsubsection{Image Style Transfer}

The aim of the image style transfer is to convert a content image into a certain stylized image.  
Currently, two types of approaches are often used for image style transfer. 

{First, when a pair of content image and a reference style image is given, the goal is to convert the content image to imitate the reference style.}
For example, the adaptive instance normalization (AdaIN) has been proposed as a simple but powerful method \cite{huang2017arbitrary} for image style transfer. 
Specifically, AdaIN layers of a network estimate means and variance of the given reference style features and use the learned parameters to adjust those of the content image. 

On the other hand, unsupervised style transfer approaches such as CycleGAN \cite{Zhu_2017_ICCV} learn target reference style distribution rather than individual style given a single image.
Unfortunately, the CycleGAN approach requires $N(N-1)$ generators to translate between $N$ domains. 
To deal with this, multi-domain image translation approaches have been proposed.
In particular, StarGANv2 \cite{choi2020stargan} introduces an advanced single generator-based framework, which transfers styles over multiple domains by training domain-specific style codes of AdaIN.

\subsubsection{Semi and Self-supervised Learning}

For medical image segmentation, a DL-based model trained with labeled dataset in a specific domain (e.g. normal CXR) is often needed to be refined for different domain dataset in semi-supervised, self-supervised or unsupervised manners \cite{tang2019XLSor, tarvainen2017mean, perone2019unsupervised, xue2020dual, orbes2019multi, li2020transformation,  chen2020unsupervised}. 
These approaches try to take advantage of learned features from a specific domain to generate pseudo-labels or distillate the learned knowledge to another domain.  

{Specifically, Tang et al. \cite{tang2019XLSor} applies a semi-supervised learning approach by generating pseudo CXRs via an image-to-image style transfer framework, and achieves improved segmentation performance by training a segmentation network with both labeled and pseudo-labeled dataset.}  

{Self-supervised learning approaches can also bring large improvements in image segmentation tasks, by promoting consistency between model outputs given a same input with different perturbations or by training auxiliary proxy tasks  \cite{xue2020dual, orbes2019multi, li2020transformation}. In general, DL models trained with auxiliary self-consistency losses are proven to achieve better generalization capability as well as better primary task performance, especially when training with abundant unlabeled dataset.}

\subsubsection{Teacher-Student Approaches}

{Teacher-student approaches can be utilized for semi or self-supervised learning framework. These methods consist of two individual networks, i.e., a student and a teacher model. 
The student model is trained in a supervised manner, as well as in a self-supervised manner which enforces the student model outputs to be consistent with outputs from the teacher model \cite{li2020transformation, perone2019unsupervised, tarvainen2017mean}.}

{Specifically, Li et al. \cite{li2020transformation} introduce a dual-teacher framework on segmentation task, which consists of two teacher models: a traditional teacher model for transferring intra-domain knowledge and an additional teacher model for transferring inter-domain knowledge. 
To leverage inter-domain dataset, images from different domain are firstly style-transferred as pseudo-images using CycleGAN \cite{Zhu_2017_ICCV}. 
Then the student network is trained to predict consistent outputs with those of the inter-domain teacher given the style-transferred images, so that the acquired inter-doamin knowledge can be integrated into the student network.}

\section{Methods}
\label{sec:main}

\subsection{Key Idea}

One of the unique features of StarGANv2 \cite{choi2020stargan} is that it can synergistically learn common features across multiple image domains via shared layers to fully utilize all different datasets, but still allow domain-specific style transfer using different style codes. 
Inspired by this idea, 
our framework adapts AdaIN layers to perform style transfer between normal and abnormal domains and further utilizes an additional style code for self-supervised learning.

Specifically, our framework categorizes the training data into three distinct groups: the segmentation mask [MASK], their matched input image domain [INTRA], and domain-shifted input images with no segmentation labels [INTER]. 
Due to the domain shift between [INTRA] and [INTER] domains, a network trained in a supervised manner using only [INTRA] dataset does not generalize well for [INTER] domain.
To mitigate this problem, we propose a single generator to perform supervised segmentation (see  Fig.~\ref{fig:scenario}(a)) as well as domain adaptation between [INTRA] and [INTER] domains using different AdaIN codes (see Fig.~\ref{fig:scenario}(b)).

Similar to the existing teacher-student approaches, we then introduce a self-consistency loss between different model outputs given learned task-specific AdaIN codes. 
The teacher network considers the indirect segmentation through domain adaptation followed by segmentation (indicated as the blue arrow in Fig.~\ref{fig:scenario}(c)), and the student network considers the direct segmentation (indicated as the red arrow in Fig.~\ref{fig:scenario}(c)).
The network is trained in a self-supervised manner that enforces the direct segmentation results to be consistent with the indirect segmentation results, 
so that unlabeled domain images can be directly segmented.
This enables knowledge distillation from learned segmentation and domain adaptation tasks to the self-supervised segmentation task.

Once the network is trained, only a single generator and pre-built AdaIN codes can be simply utilized at the inference phase, which makes the proposed method more practical.

\subsection{Overall Framework}

\begin{figure*}[bth!]
\centering
\includegraphics[width=12cm]{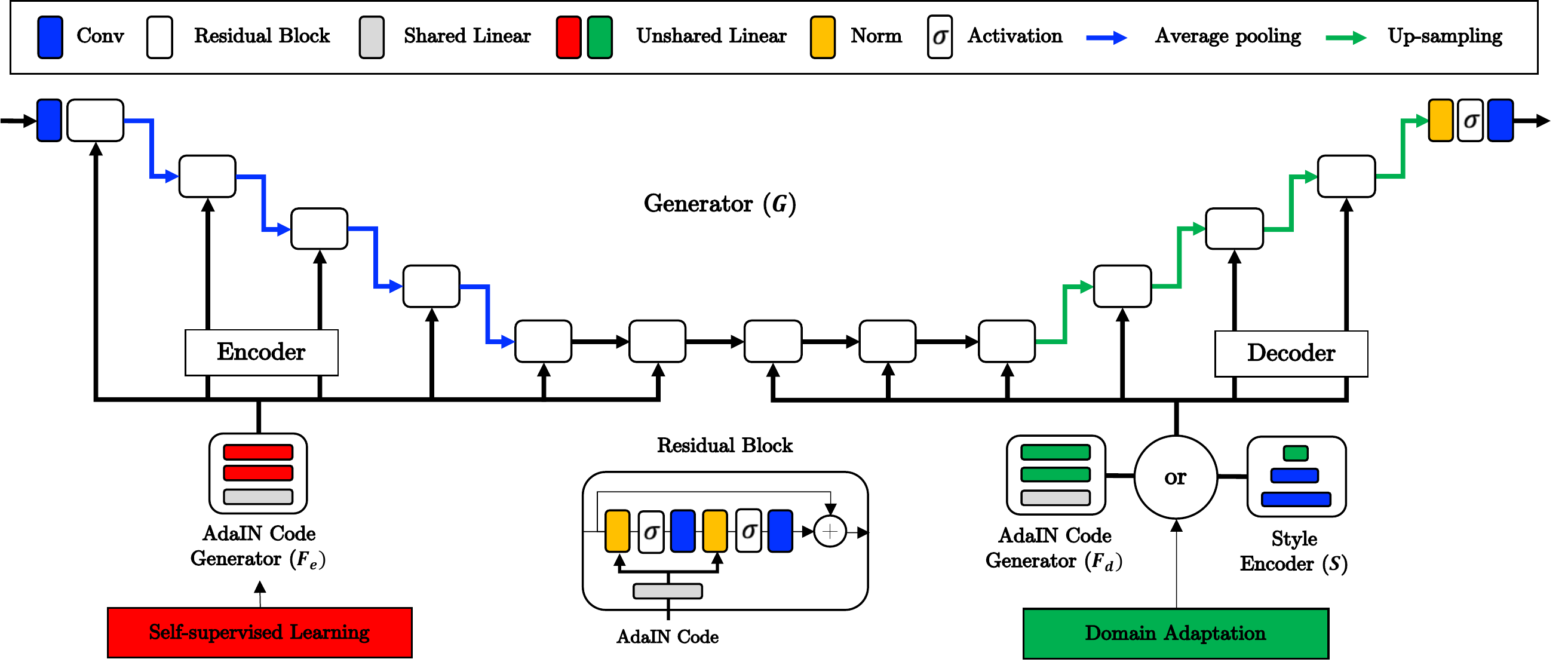}
\caption{The architecture of generator connected to two AdaIN code generators and a style encoder. The codes generated from either AdaIN code generators or the style encoder are connected to AdaIN layers of each residual block.}
\label{fig:network_overview}
\end{figure*}

Overall architecture of our network is shown in Fig. \ref{fig:network_overview},
which is composed of a single generator $G$, AdaIN code generators for the encoder and decoder, $F_e$ and $F_d$, respectively, a style encoder $S$, and a multi-head discriminator $D$.

The generator $G$ is composed of encoder and decoder modules. 
Specifically, the encoder part is composed of four downsampling residual blocks and two intermediate residual blocks. The decoder part  is composed of two intermediate residual blocks and four up-sampling residual blocks. 
Each residual block is composed of AdaIN layers, activation layers, and convolution layers. 
All the AdaIN layers are connected to the code generators $F_e$ and $F_d$, and the style encoder $S$ is also connected to the decoder module.
Detailed network specification is provided in Supplementary Section \ref{sp:architecture}.

One of key ideas of our framework is introducing independent code generators $F_e$ and $F_d$ to each encoder and decoder module. 
Thanks to the two separate code generators, the generator $G$ can perform segmentation, domain adaptation and self-supervised learning tasks, by simply changing combinations of the AdaIN codes, as shown in Table \ref{table:codes}.

\begin{table}[h!]
  \caption{AdaIN codes combination for three different tasks.}
  \label{table:codes}
  \centering
  \scalebox{0.8}{
\begin{tabular}{M{2.4cm}|M{1.8cm}M{1.8cm}M{2.4cm}M{2.5cm}M{3.5cm}} 
\hline  
  \multirow{2}{*}{\bf{  AdaIN code }}& \multirow{2}{*}{\bf{[Source]}} & \multirow{2}{*}{\bf{[Target]}} &  $\bf{F_e}$ &  $\bf{F_d}$   &  \multirow{2}{*}{ \bf{Task}}  \\ 
~ &  & & {  (mean, var)  }  &  { (mean, var)  } &  ~ \\
  \hline
  \hline
$a_{seg}$&  [INTRA] & [MASK]  & (0, 1) & (0, 1) & {segmentation} \\
$a_{seg}'$&  [INTRA] & [MASK]  & (0, 1) & learnable & {dummy segmentation} \\
  \hline
$a_{da}^{X}$& [INTER] & [INTRA]  & (0, 1) &  learnable  & domain adaptation \\
$a_{da}^{Y}$&  [INTRA] & [INTER]  & (0, 1) &  learnable  & domain adaptation \\
  \hline
$a_{self}$ &  [INTER] & [MASK]    & learnable & (0, 1) &self-supervised \\
\hline  
  \end{tabular}  }
\end{table}


Specifically, let $\Xc,\Yc$ and $\Zc$ refer to the [INTRA], [INTER] and [MASK] domains associated with the probability distribution $P_X$, $P_Y$ and $P_Z$.
Then, our generator is defined by
\begin{align}
v=G(u, a),\quad a := (F_e, F_d)
\end{align}
where $u$ is the input image either in $\Xc$ or $\Yc$, {$a\in \{a_{seg}, a_{seg}',a_{da}^{X}, a_{da}^{Y}, a_{self}\}$ refers the AdaIN code, as shown in Table~\ref{table:codes}, and $F_e$ and $F_d$ indicate the code generators for the encoder and the decoder, respectively.
Given the task-specific code $a$, AdaIN layers of the generator $G$ efficiently shift weight distribution to desirable style distribution, by adjusting means and variances.
Detailed style transfer mechanism of AdaIN is provided in Supplementary Section \ref{sp:adain}. 
Hence, the generator $G$ can generate output $v$ either in $\Xc,\Yc$, or $\Zc$ domain, depending on different codes combination. 

The style encoder is introduced by StyleGANv2 to impose an additional constraint, so that the learned AdaIN codes should reflect style of the given reference images \cite{choi2020stargan}.
In our framework, the style encoder $S$ encodes the generated output into a code for imposing code-level cycle-consistency. 
Another role of the style encoder $S$ is to generate reference-guided codes given reference images.
By alternatively generating codes using the style encoder $S$ or the AdaIN code generator $F_d$, as illustrated as the $or$ module in Fig. \ref{fig:network_overview}, the learned codes can be regularized to reflect the reference styles.

\begin{figure}[hbt!]
\centering
\includegraphics[width=9cm]{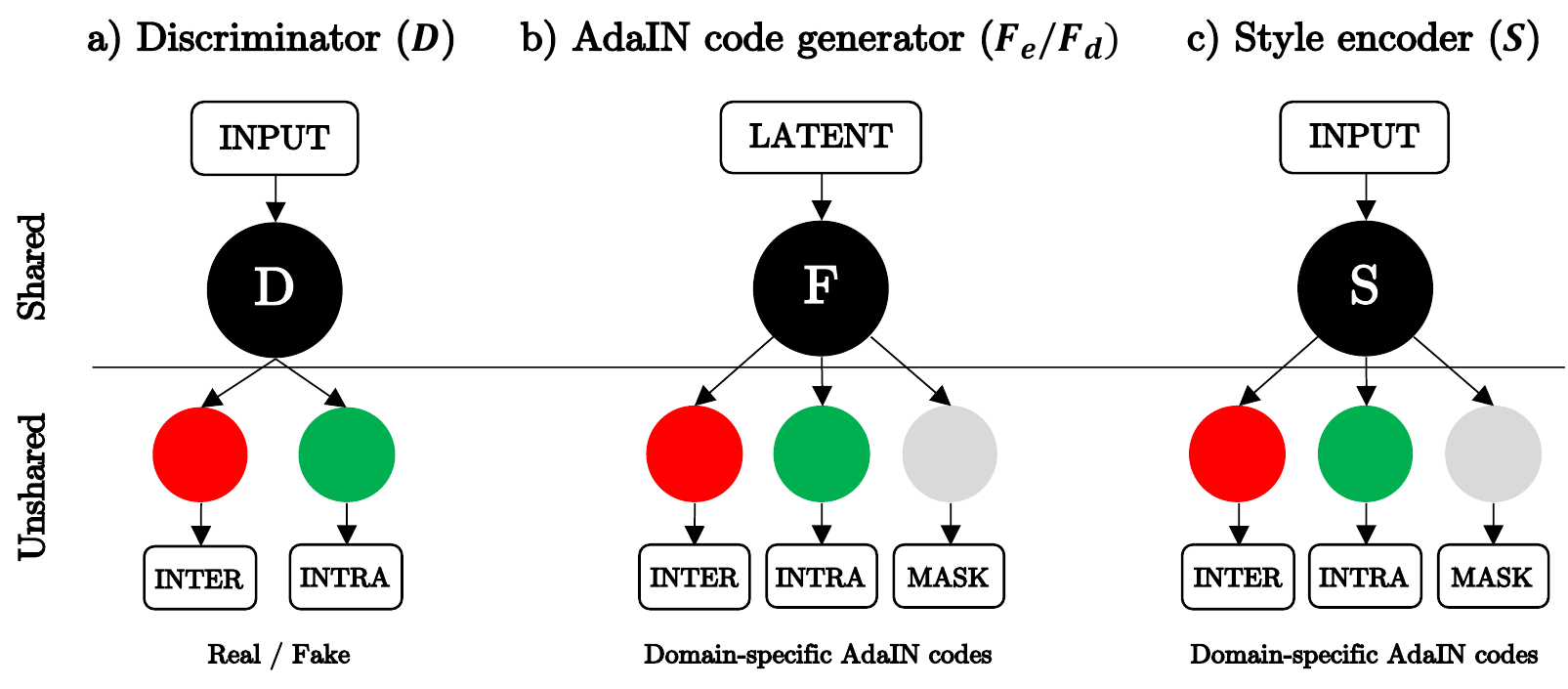}
\caption{Multi-head structure of a) Discriminator, b) AdaIN code generator, and c) Style encoder. Each module is composed of shared  layers and domain-specific layers.}
\label{fig:multihead}
\end{figure}

{The discriminator  $D$ is composed of shared convolution layers followed by multi-headed unshared convolution layers for each image domain, as shown in Fig. \ref{fig:multihead}(a). In the discriminator, the input image can be classified as 1 or 0 for each domain separately, where 1 indicates real and 0 indicates fake.} 
{The AdaIN code generator and the style encoder are also composed of shared layers followed by domain-specific unshared layers, as shown in Fig. \ref{fig:multihead}(b) and (c). }
Thanks to the existence of the shared layers, learned features from a specific domain can be shared with other domains, improving overall performance of each module.

In the following, we provide more detailed description how this network can be trained.

\subsection{Neural Network Training}

Our training losses are extended from traditional style transfer framework, with specific 
modification to include the segmentation and the self-supervised learning tasks. The details are as follows.

\subsubsection{Supervised Segmentation}

This part is a unique contribution of our work compared to traditional style transfer methods.
Fig. \ref{fig:scenario}(a) shows the supervised segmentation, which can be considered as  conversion from $\Xc$ to $\Zc$.
In this case, the generator is trained by the following: 
\begin{align}
\min_{G,F_d,S} &~ \lambda_{seg}\ell_{seg}(G) \label{eq:seg} 
+\lambda_{style} \ell_{style}(G, F_d, S),
\end{align}
where $\lambda_{seg}$ and $\lambda_{style}$ are hyper-parameters, and the segmentation loss $\ell_{seg}$ is defined by the cross-entropy loss between generated output and its matched label: 
\begin{align}
\ell_{seg}(G) = 
-\mathbb{E}_{x\sim P_X} \left[z_i \log p_i(G(x, a_{seg}))\right],
\end{align}
where $z_i$ denotes the $i$-th pixel of the ground truth segmentation mask $z\in \Zc$ with respect to the input image $x \in \Xc$,
$p_i(G)$ denotes the softmax probability function of the $i$-th pixel in the generated image, and $G(x, a_{seg})$ denotes the supervised segmentation task output.

Once a segmentation result is generated, the style encoder $S$ encodes the generated image to be consistent with the dummy AdaIN code $a_{seg}'$.
This can be achieved by using the following style loss:
\begin{align}
&\ell_{style}(G, F_d, S) 
= \mathbb{E}_{x\sim P_X} 
\left[ \lVert {{a}'_{seg}}  - {S}(G(x, {a}_{{seg}}))
 \rVert_{1} \right],
\label{eq:style}
\end{align}
where ${{a}'_{seg}}$ can be either generated by $F_d(z)$ or $S(z)$, given segmentation mask $z\in \Zc$.
Although this code is not used for segmentation directly, the generation of this dummy AdaIN code turns out to be important to train the shared layers in the AdaIN code generator and the style encoder. We analyzed contribution of different losses for the supervised segmentation task, as analyzed in Supplementary Section \ref{sp:ablation}. 
%

\subsubsection{Domain Adaptation}

Fig. \ref{fig:scenario}(b) shows the training scheme for the domain adaptation between $\Xc$ and $\Yc$. The training of domain adaptation solves the following optimization problem:
\begin{align}
\min_{G, F_e,F_d,S}  \max_{D} \ell_{da}(G,F_e,F_d, S, D) . \label{eq:da}
\end{align}

The role of Equation \ref{eq:da} is to train the generator $G$ to synthesize style-transfered images given domain-specific AdaIN codes $a_{da}$, while fooling the discriminator $D$. As this step basically follows traditional style transfer methods, the detailed domain adaptation loss is deferred to Supplementary Section \ref{sp:da}.

\subsubsection{Self-supervised Learning}

This part is another unique contribution of our work. 
The goal of the self-supervised learning is to directly transfer an unlabeled image in $\Yc$ to segmentation mask $\Zc$, illustrated as the red arrow in Fig. \ref{fig:scenario}(c).

Specifically, since [INTER] domain $\Yc$ lacks segmentation mask, knowledge learned from both the supervised learning and domain adaptation need to be distilled. 
Thus, our contribution comes from  introducing  novel constraints: (1) the direct segmentation outputs trained in a self-supervised manner, illustrated as the red arrow in Fig. \ref{fig:scenario}(c) should be consistent with indirect segmentation outputs, illustrated as the blue arrow in Fig. \ref{fig:scenario}(c). 
(2) At the inference phase, it is often difficult to know which domain the input comes from. 
Therefore, a single AdaIN code should deal with both [INTRA] and [INTER] domain image segmentation. 
This leads to the following self-consistency loss:
\begin{align}
& \ell_{self}(G,F_e) \label{eq:self} \\
&= \lambda_{inter} \ell_{inter}(G, F_e) + \lambda_{intra} \ell_{intra}(G, F_e)  \\
&= \mathbb{E}_{y\sim P_Y}
 \left[ \lVert  G^\prime(G^\prime(y, {a}_{da}^{\Tc}),  {a}_{seg}) 
-G(y, {a}_{self})
 \rVert_{1} \right] \label{eq:self_inter}\\
 &+\mathbb{E}_{x\sim P_X}
 \left[ \lVert  G(x, {a}_{self})
-G^\prime(x, {a}_{seg})
 \rVert_{1} \right] \label{eq:self_intra},
\end{align}
where $G^\prime$ indicates the frozen generator $G$, $\ell_{inter}$ and $\ell_{intra}$ denote inter-domain and intra-domain self-consistency loss, respectively, and $\lambda_{inter}$ and $\lambda_{intra}$ are hyper-parameters for each. The role of Eq.~\eqref{eq:self_inter} is for emposing the constraint (1), and Eq.~\eqref{eq:self_intra} is for emposing the constraint (2).

In fact, this procedure can be regarded as a teacher-student approach.
The indirect path is a teacher network that guides the training procedure of the direct path, which regards the student network. 
In contrast to the existing  teacher-student approaches \cite{tarvainen2017mean, li2020transformation, perone2019unsupervised}, our approach does not need two separate networks: instead, the single generator with different AdaIN codes combinations can be served as the teacher or the student network, which is another big advantage of our method.


\section{Experiments}
\label{sec:methods}

\subsection{Experimental settings}

\subsubsection{Dataset}

For training the supervised segmentation task, normal CXRs were acquired from JSRT dataset \cite{shi2000jsrt} with their paired lung segmentation labels from SCR dataset \cite{van2006scr}.
For training domain adaptation task, we collected pneumonia CXRs from RSNA \cite{desai2020chest} and Cohen dataset \cite{cohen2020covid}. Detailed dataset information is described in Table \ref{table:xraydatabase}.

\begin{table*}[h!]
	\caption{ Chest X-ray dataset resources.}
	\label{table:xraydatabase}
	\centering
	\scalebox{0.70}{
		\begin{tabular}{l|ccccc|cccc|cc}
			\hline  
			\multirow{2}{*}{\textbf{Domain}} & \multirow{2}{*}{\textbf{Dataset}} & \multirow{2}{*}{\textbf{Label}}   & \multirow{2}{*}{\textbf{Disease class}} & \multirow{2}{*}{\textbf{Bit}} & \multirow{2}{*}{\textbf{View}}  & \multirow{2}{*}{\textbf{Total}}   & \multirow{2}{*}{\textbf{Train}} & \multirow{2}{*}{\textbf{Val}} & \multirow{2}{*}{\textbf{Test}} & \multicolumn{2}{c}{\textbf{DS level}} \\
			\cline{11-12}
			& &  & & & &  & & &  & {\emph{Weak}} & {\emph{Harsh}} \\
			\hline  
			\hline
			Labeled train set	& JSRT & O  	& Normal, Nodule 			& 12  & PA 	& 247  	& 178 & 20  & 49 & - \\ 
			\hline
			\multirow{3}{*}{Unlabeled train set  } & RSNA & -  	& PN (COVID-19)		& 10  & AP   	& 218 	& 218 & - & - & - \\ ~
			~ & \multirow{2}{*}{Cohen et al.} & \multirow{2}{*}{-}	&  {PN (COVID-19,}   	& \multirow{2}{*}{8} & \multirow{2}{*}{PA, AP} & \multirow{2}{*}{680}  & \multirow{2}{*}{640} & \multirow{2}{*}{-} & \multirow{2}{*}{40} & \multirow{2}{*}{-}  \\
			~						& ~ 	& ~			&  {Viral, Bacterial, TB)} &  &  &  &  &  &  & \\
			\hline
			\multirow{4}{*}{External testset}  & {NLM} 	& O		& Normal 				& 8  	& PA 	& 80  	& - & - &80 & {80} & {80} \\ 
			~ 						& {BIMCV-13} & O  & PN (COVID-19)  	& 16 	& PA, AP 	& 13  	& - & - & 13 & {13} & {13}  \\
			~ 						& {BIMCV} & - 	 	& PN (COVID-19)  	& 16 	& AP 	& 374  	& - & - & 374 & - & - \\
			~						& {BRIXIA} & - 		& PN (COVID-19)  	& 16 	& AP 	& 2384  	& - & - & 2384 & - & - \\
			\hline  
			\multicolumn{11}{l}{\textit{Note:} PN, pneumonia; TB, tuberculosis; DS level, distribution shift level.}\\
	\end{tabular}  }
\end{table*}

{To test the proposed network performance on external datasets, we utilized three external resources. For normal CXR segmentation evaluation, NLM dataset with paired lung labels were utilized \cite{jaeger2014two}. For abnormal CXR segmentation evaluation, BIMCV dataset \cite{vaya2020bimcv} and BRIXIA dataset \cite{signoroni2021bs} were utilized. 
Besides, additional 13 CXRs from the BIMCV dataset (indicated as BIMCV-13), with labeled consolidation or ground glass opacities features by radiologists, were utilized for quantitative evaluation of abnormal CXRs segmentation.}

{For further analyzing the model performance on domain-shifted conditions, external labeled dataset were prepared with three different levels of modulation, defined as distribution shift level. \emph{None}-level indicates original inputs. \emph{Weak}- and \emph{Harsh}-level indicate intensity and contrast modulated inputs with random scaling factors within $\pm$30\% and $\pm$60\% range, followed by addition of Gaussian noise with standard deviation of 0.5 and 1, respectively.}

All the input CXRs and labels are resized to $256 \times 256$. We did not perform any pre-processing or data augmentation except for normalization of pixel intensity range to  [-1,0, 1.0].

\subsubsection{Implementation Details}

The proposed network was trained by feeding input images from a pair of two randomly chosen domains: one for the source domain and the other for the target domain. For example, if a domain pair composed of [INTER] and [INTRA] domains fed into the network, the network was trained for the domain adaptation task. When a domain pair, composed of [INTRA] as source and [MASK] as target domain, fed into the network, the network was trained for the supervised segmentation task. 
For self-supervised learning, an image from [INTER] domain, as well as from [INTRA], was fed as the source domain to output the segmentation mask. 
Implementation details for training the proposed network are provided in Supplementary Section \ref{sp:mplementation}. 
 
For the domain adaptation task, we utilized {CycleGAN \cite{Zhu_2017_ICCV}}, MUNIT \cite{huang2018multimodal} and StarGANv2 \cite{choi2020stargan} as baseline models for comparative studies. For the segmentation task, we utilized U-Net \cite{ronneberger2015unet} as a baseline model.  All the baseline models were trained with identical conditions to that of the proposed model.
To evaluate unified performance of the domain adaptation task and the segmentation task, we utilized available models for performing abnormal CXR segmentation, i.e., XLSor \cite{tang2019XLSor} and {lungVAE \cite{selvan2020lung}.} 
 Implementation details for all the comparative models are provided in Supplementary Section \ref{sp:compare}. 

\subsubsection{Evaluation Metric}

For normal CXRs, quantitative segmentation performance of both lungs was evaluated using dice similarity score (Dice) index.
The abnormal CXR segmentation performance was evaluated quantitatively using true positive ratio (TPR) of the annotated abnormalities labels.}
Moreover, for unlabeled abnormal dataset, domain adaptation and segmentation performance were qualitatively evaluated based on generation of expected lung structure covered with highly-consolidated regions.

\subsection{Results}
\label{sec:result}

\subsubsection{Unified Domain Adaptation and Segmentation}
\begin{figure*}[h!]
\centering
\includegraphics[width=12cm]{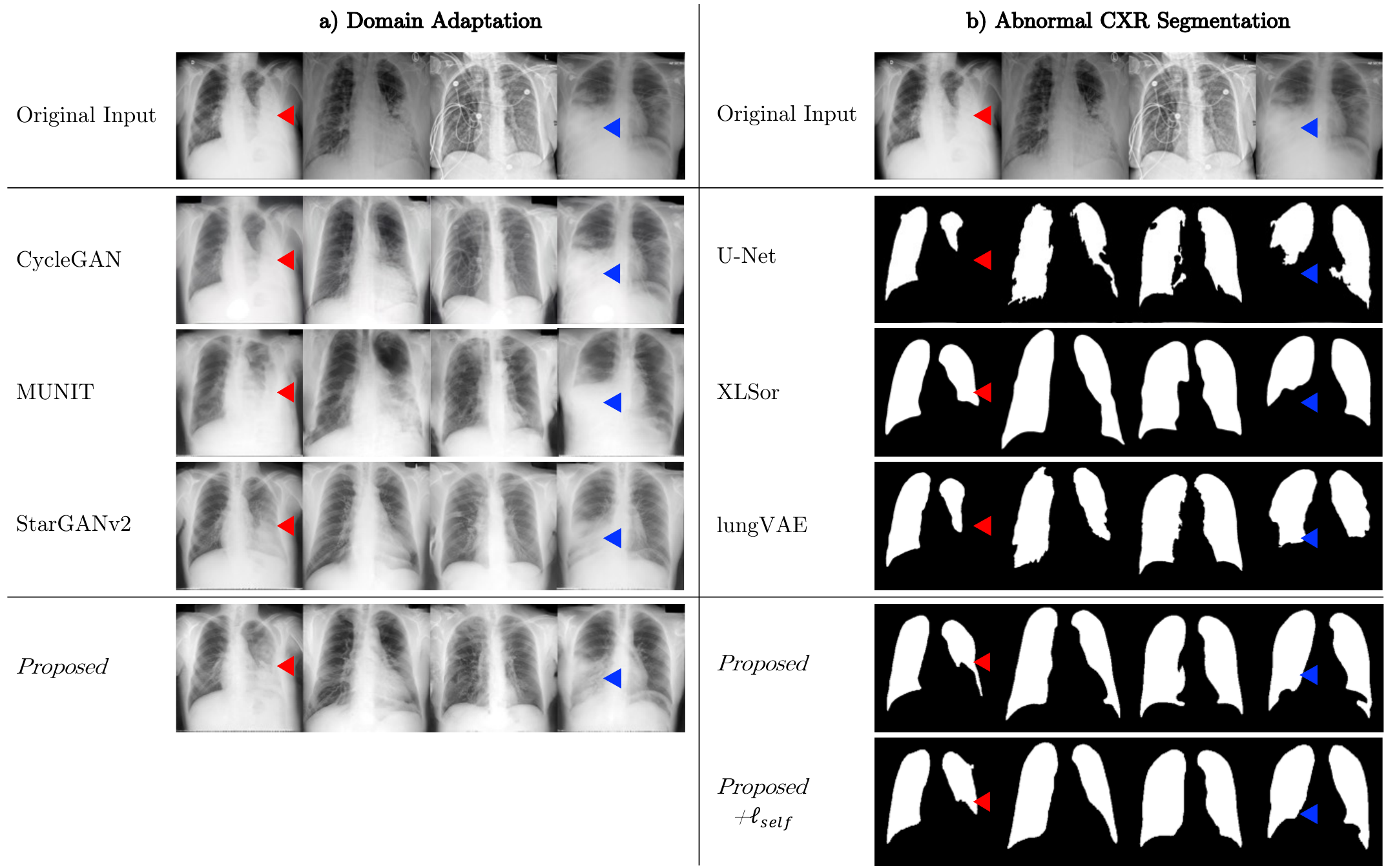}
\caption{The network performance on the internal test set. a) Domain adaptation performance comparison with traditional style transfer methods. b) Segmentation performance comparison. Red and blue triangles indicate highly consolidated lung regions in CXRs.}
\label{fig:results_proposed}
\end{figure*}

The unified performance was evaluated on the internal test set. {We defined our model trained with segmentation loss (Eq.~\eqref{eq:seg}) and domain adaptation loss (Eq.~\eqref{eq:da}) as \emph{Proposed}, and the model trained with additional self-consistency loss (Eq.~\eqref{eq:self}) as \emph{Proposed+$\ell_{self}$}. As shown in Fig. \ref{fig:results_proposed}(a), compared to CycleGAN and MUNIT, \emph{Proposed} model utilizing StarGANv2 framework successfully transferred highly consolidated regions in abnormal CXRs into normal lungs.}

Abnormal CXR segmentation results are presented in Fig. \ref{fig:results_proposed}(b). 
All the comparative models failed to segment highly consolidated lung regions, indicated as red and blue triangles. 
Note that \emph{Proposed} and \emph{Proposed+$\ell_{self}$} models were the only methods that successfully segmented abnormal lungs as like normal lungs.

\subsubsection{Quantitative Evaluation on Domain-shifted External Dataset} 
{For quantitative evaluation, we utilized labeled dataset for both normal and abnormal CXRs (NLM and BIMCV-13, respectively).
To verify that the proposed methods could still retain segmentation performance on domain-shifted dataset, we further tested the model performance on distribution modulated inputs (\emph{Weak}- and \emph{Harsh}-level) for both normal and abnormal dataset.}

\begin{figure}[h!]
\centering
\includegraphics[width=12cm]{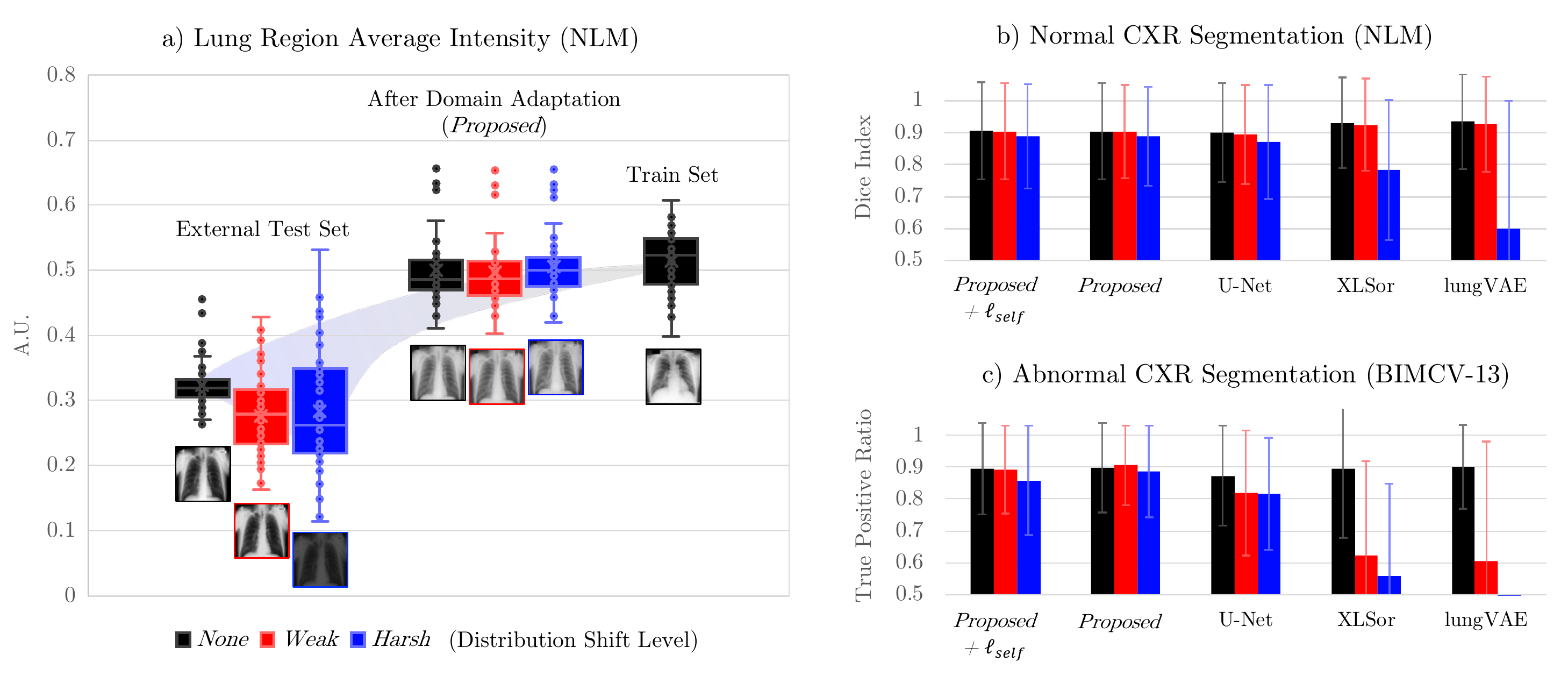}
\caption{Model performance on domain-shifted inputs. a) Box plot of average lung region intensity of normal CXRs with different distribution shift levels. Bar plots of segmentation performance on b) Normal, and c) Abnormal CXR dataset. Black, red and blue bal plots indicate \emph{None}-,  \emph{Weak}- and \emph{Harsh}-level dataset, respectively.}
\label{fig:results_table}
\end{figure}

As illustrated in Fig. \ref{fig:results_table}(a), \emph{Proposed} model successfully adapted shifted distribution of lung area intensity to be similar to the train set distribution.  
Thanks to the successful domain adaptation performance, both \emph{Proposed} and \emph{Proposed+$\ell_{self}$} models maintained robust segmentation performance compared to other models,
as distribution gap increases (see blue bar plots of Fig. \ref{fig:results_table}(b) and (c)).

{Specifically, as reported in Table \ref{table:compare}, for the abnormal CXR segmentation task, all the models showed promising performance by achieving TPR of around 0.90 for the original (\emph{None}-level) inputs, illustrated as black bar plots in Fig. \ref{fig:results_table}(c). However, XLSor and lungVAE performance drastically dropped to around 0.60 for \emph{Weak}-level inputs, while \emph{Proposed} model performance rather improved, as illustrated in red bar plots. 
For \emph{Harsh}-level inputs, lung structures were only correctly segmented by \emph{Proposed} and \emph{Proposed+$\ell_{self}$} models with above 0.85 of TPR, as illustrated in blue bar plots. }

\begin{table}[h!]
  \caption{Segmentation performance on external test set. }
  \label{table:compare}
  \centering
  \scalebox{0.66}{
\begin{tabular}{l|M{2.3cm}M{2.3cm}M{2.3cm}|M{2.3cm}M{2.3cm}M{2.3cm}} 
\hline  
  \multirow{3}{*}{\textbf{Method}} & \multicolumn{3}{c|}{\textbf{Normal CXR (Dice Index)}} & \multicolumn{3}{c}{\textbf{Abnormal CXR (True Positive Ratio)}} \\ 
    \cline{2-7}
    & \multicolumn{3}{c|}{Distribution Shift Level} & \multicolumn{3}{c}{Distribution Shift Level} \\ 
  & \emph{None} & \emph{Weak} & \emph{Harsh} & \emph{None} & \emph{Weak} & \emph{Harsh} \\
  \hline
    \hline
\textbf{SS}&&&&&&\\
U-Net \cite{ronneberger2015unet} 	&{0.90} $\pm$ {0.15} 			& {0.89} $\pm$ 0.16		& {0.87} $\pm$ 0.18			& 0.87 $\pm$ 0.16 			& 0.82 $\pm$ 0.20 			& 0.82 $\pm$ 0.18 \\
 \hline
\textbf{DA+SS}&&&&&& \\
CycleGAN \cite{Zhu_2017_ICCV}+U-Net 	&{0.89} $\pm$ {0.17} 			& {0.89} $\pm$ 0.17		& {0.86} $\pm$ 0.18			& 0.88 $\pm$ 0.17 			& 0.84 $\pm$ 0.22			& 0.85 $\pm$ 0.17 \\
StarGANv2 \cite{choi2020stargan}+U-Net &\textbf{0.90} $\pm$ {0.15} 		& \textbf{0.90} $\pm$ 0.15		& {0.88} $\pm$ 0.15			& \textbf{0.90} $\pm$ 0.13 	& 0.90 $\pm$ 0.12 			& 0.88 $\pm$ 0.16 \\
\emph{Proposed}					&\textbf{0.90} $\pm$ 0.01 			& \textbf{0.90} $\pm$ {0.14}	& \textbf{0.89} $\pm$ {0.15}	& \textbf{0.90} $\pm$ 0.14 	 & \textbf{0.91} $\pm$ {0.12} 	& \textbf{0.89} $\pm$ {0.14} \\
\hline
\textbf{UDS/Self}&&&&&& \\
XLSor \cite{tang2019XLSor} 	&{0.93} $\pm$ {0.14} 		& {0.92} $\pm$ 0.15 			& {0.78} $\pm$ 0.22 			& \textbf{0.90} $\pm$ 0.22 	& 0.62 $\pm$ 0.30 			& 0.56 $\pm$ 0.29 \\
lungVAE  \cite{selvan2020lung}	& \textbf{0.94} $\pm$ {0.15}	&\textbf{0.93} $\pm$ {0.15} 	& {0.60} $\pm$ 0.40 			& \textbf{0.90} $\pm$ {0.13}	& 0.61 $\pm$ 0.38 			& 0.10 $\pm$ 0.26 \\
\emph{Proposed+$\ell_{self}$ } & 0.91 $\pm$ 0.15			& {0.90} $\pm$ {0.15} 		& \textbf{0.89} $\pm$ {0.16} 	& \textbf{0.90} $\pm$ 0.14 	& \textbf{0.89} $\pm$ {0.14} 	&\textbf{0.86} $\pm$ {0.17}\\
\hline  
  \multicolumn{7}{l}{\textit{Note:} SS, supervised segmentation; DA, domain adaptation; UDS, unified DA+SS; Self, self-supervised segmentation.}\\
  \end{tabular}  }
\end{table}

\begin{figure}[h!]
\centering
\includegraphics[width=12cm]{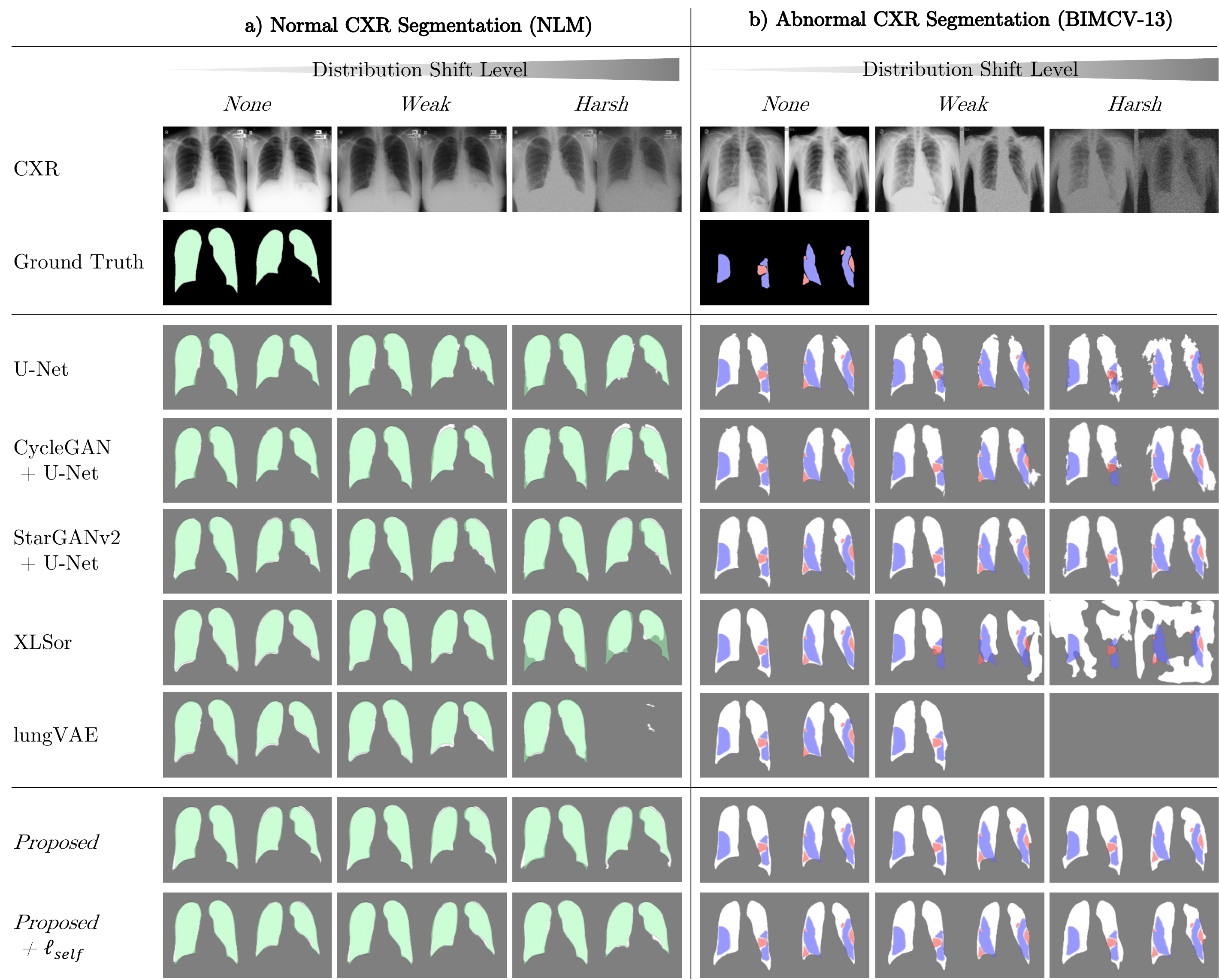}
\caption{Qualitative segmentation performance for a) Normal CXRs, and b) Abnormal CXRs. Green ground truth masks indicates lung labels, red and blue masks indicate consolidation and ground-glass opacity labels, respectively}
\label{fig:result_dshift}
\end{figure}

Corresponding segmentation contours are also presented in Fig. \ref{fig:result_dshift}. The qualitative segmentation performance was surprising since artificial perturbation, which seemed tolerable for human eye, brought strong performance degradation to the existing DL-based algorithms. 
Specifically, for abnormal CXR segmentation, all the comparative UDS/Self methods (XLSor and lungVAE) failed to be generalized to several out-of-distribution cases from \emph{Harsh}-level inputs (see Fig. \ref{fig:result_dshift}(b)). The UDS/Self methods have no additional domain adaptation process, but the networks themselves are trained with augmented data distribution, e.g., images added with random noise or pseudo-pneumonia images. The degraded performance indicates that the above data augmentation techniques are still limited to be generalized to way-shifted data distribution.
U-Net with or without style-transfer based pre-processing (CycleGAN or StarGANv2), rather endured harsh-distribution shift, however, lung contours showed irregular shapes with rough boundary.
On the other hand, \emph{Proposed} and \emph{Proposed+$\ell_{self}$} models showed stable segmentation performance on majority of domain-shifted cases. 
In particular, \emph{Proposed+$\ell_{self}$} model showed most promising performance, with the least over-segmentation artifact for both normal and abnormal datasets.

\subsubsection{Qualitative Evaluation on COVID-19 Dataset} 

\begin{figure*}[h!]
\centering
\includegraphics[width=12cm]{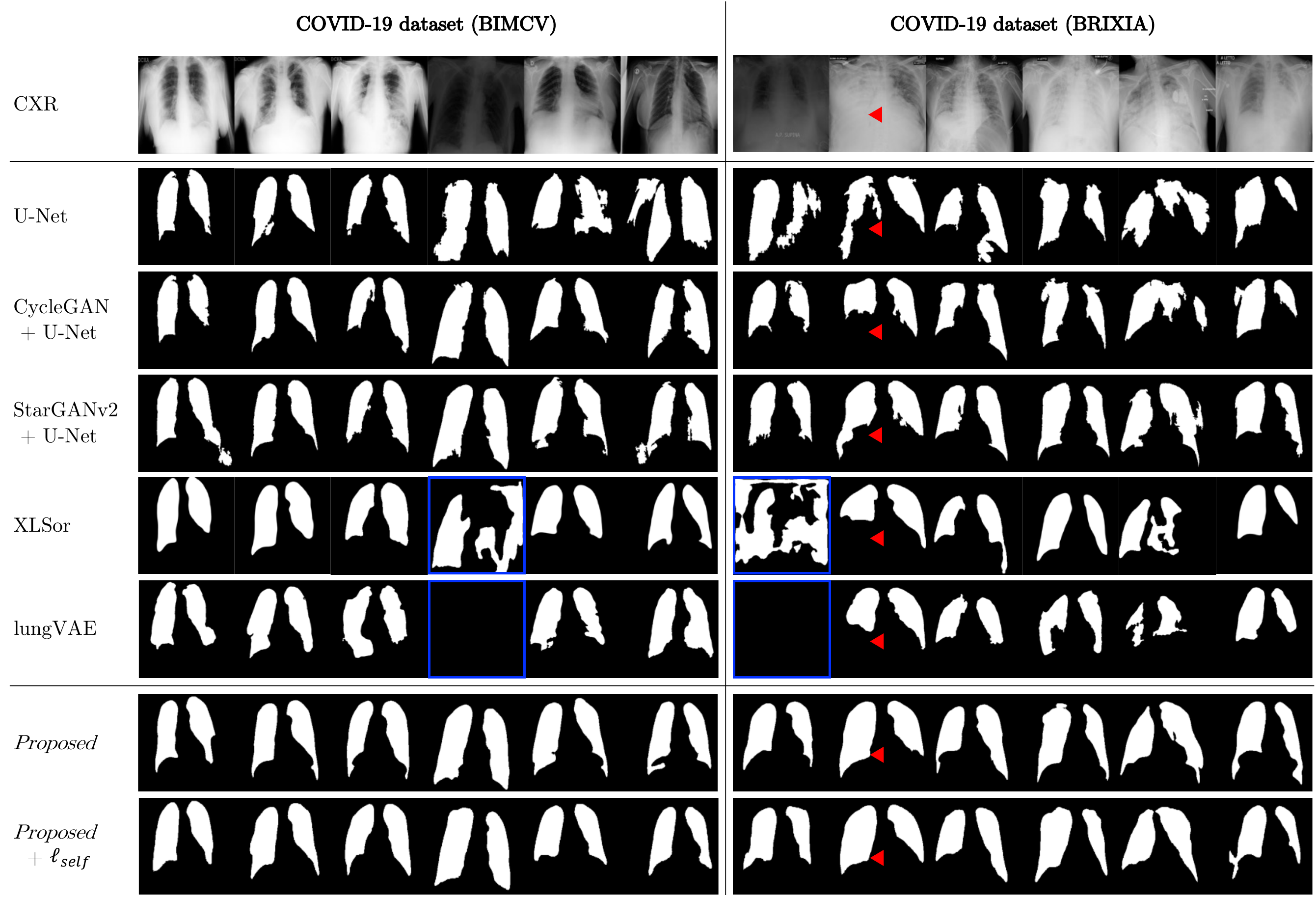}
\caption{Abnormal CXR segmentation results on external COVID-19 dataset. Red triangles indicate highly consolidated lung regions, and blue boxes indicate segmentation results which fail to be generalized to out-of-distribution data.}
\label{fig:result_compare}
\end{figure*}

{For evaluating the model performance on real-world dataset, we utilized COVID-19 pneumonia dataset (BIMCV and BRIXIA), which are obtained from more than 12 hospitals. }
Fig. \ref{fig:result_compare} presents qualitative results of abnormal CXR segmentation on COVID-19 pneumonia dataset. Representative cases were randomly selected from each dataset.

The external COVID-19 dataset showed varied intensity and noise distribution. {Comparative models mostly failed to segment regular lung shapes. Specifically, for highly consolidated regions indicated as red triangles, all the existing models were suffered from under-segmentation artifacts. For several cases, XLSor and lungVAE totally failed to be generalized to domain-shift issues (blue boxes). \emph{Proposed} and \emph{Proposed+$\ell_{self}$} models showed reliable segmentation performance without severe under-segmentation or over-segmentation artifacts.}

%
%
%

\subsubsection{Error Analysis}

We further analyzed typical error cases, which failed to be segmented, and the error cases were grouped into three categories. The representative error cases selected from each category are shown in Supplementary Section \ref{sp:error}.

\section{Conclusions}
\label{sec:conclusion}

We present a novel framework, which can perform segmentation, domain adaptation and self-supervised learning tasks within a single generator in a cost-effective manner (Supplementary Section \ref{sp:costs}). The proposed network can fully leverage knowledge learned from each task by utilizing shared network parameters, thus the model performance can be synergistically improved via knowledge distillation between multiple tasks, so that achieves SOTA performance on the unsupervised abnormal CXR segmentation task. The experimental results demonstrate that the proposed unified framework can solve domain shift issues with great generalizability, even on dataset with way-shifted distribution. Last but not the least, the proposed model does not need any pre-processing techniques but shows superior domain adaptation performance, which presents a promising direction to solve the generalization problem of DL-based segmentation methods.

\setcounter{secnumdepth}{0}
\subsection*{Acknowledgement}
This research was supported by the Korea Medical Device Development Fund grant funded by the Korea government (the Ministry of Science and ICT, the Ministry of Trade, Industry and Energy, the Ministry of Health \& Welfare, the Ministry of Food and Drug Safety) (Project Number: 1711137899, KMDF\_PR\_20 200901\_0015), the MSIT(Ministry of Science and ICT), Korea, under the ITRC (Information Technology Research Center) support program(IITP-2022-2020-0-01461) supervised by the IITP(Institute for Information \& communications Technology Planning \& Evaluation), the National Research Foundation of Korea under Grant NRF-2020R1A2B5B03001980, and the Field-oriented Technology Development Project for Customs Administration through National Research Foundation of Korea(NRF) funded by the Ministry of Science \& ICT and Korea Customs Service(NRF-2021M3I1A1097938)
\clearpage

\bibliographystyle{splncs04}
\bibliography{seg_eccv_oyj}
\clearpage

\beginsupplement
\setcounter{page}{1}
\renewcommand{\figurename}{Supplementary Figure}
\renewcommand{\tablename}{Supplementary Table}

\setcounter{secnumdepth}{1}
\setcounter{section}{0}

\title{CXR Segmentation by AdaIN-based Domain Adaptation and Knowledge Distillation: Supplementary Materials}

\titlerunning{CXR Segmentation by AdaIN-based DA and KD}
%
\author{Yujin Oh\orcidlink{0000-0003-4319-8435} \and
	Jong Chul Ye\orcidlink{0000-0001-9763-9609} }

\institute{Kim Jaechul Graduate School of AI, Korea Advanced Institute of Science and Technology (KAIST), Daejeon, South Korea
	\email{jong.ye@kaist.ac.kr}}

\authorrunning{Y. Oh et al.}

\maketitle

\section{Applications for DL-based Automatic CXR Analysis}
\label{sp:applications}

Current deep learning (DL)-based CXR analysis tasks not just diagnose disease but provide explainable results such like saliency map or quantified severity level to assist clinicians \cite{lakhani2017deep, hou2021explainable, park2021vision, frid2021covid}.
Accordingly, we further investigated applications of the proposed method using COVID-19 pneumonia CXR dataset \cite{vaya2020bimcv}.  
For generating saliency map, we referenced a public source code \footnote{\url{https://github.com/priyavrat-misra/xrays-and-gradcam.git}}.
In addition, for severity quantification, we followed array-based methods \cite{park2021vision, toussie2020clinical}. 

\begin{figure}[h!]
	\centering
	\includegraphics[width=12cm]{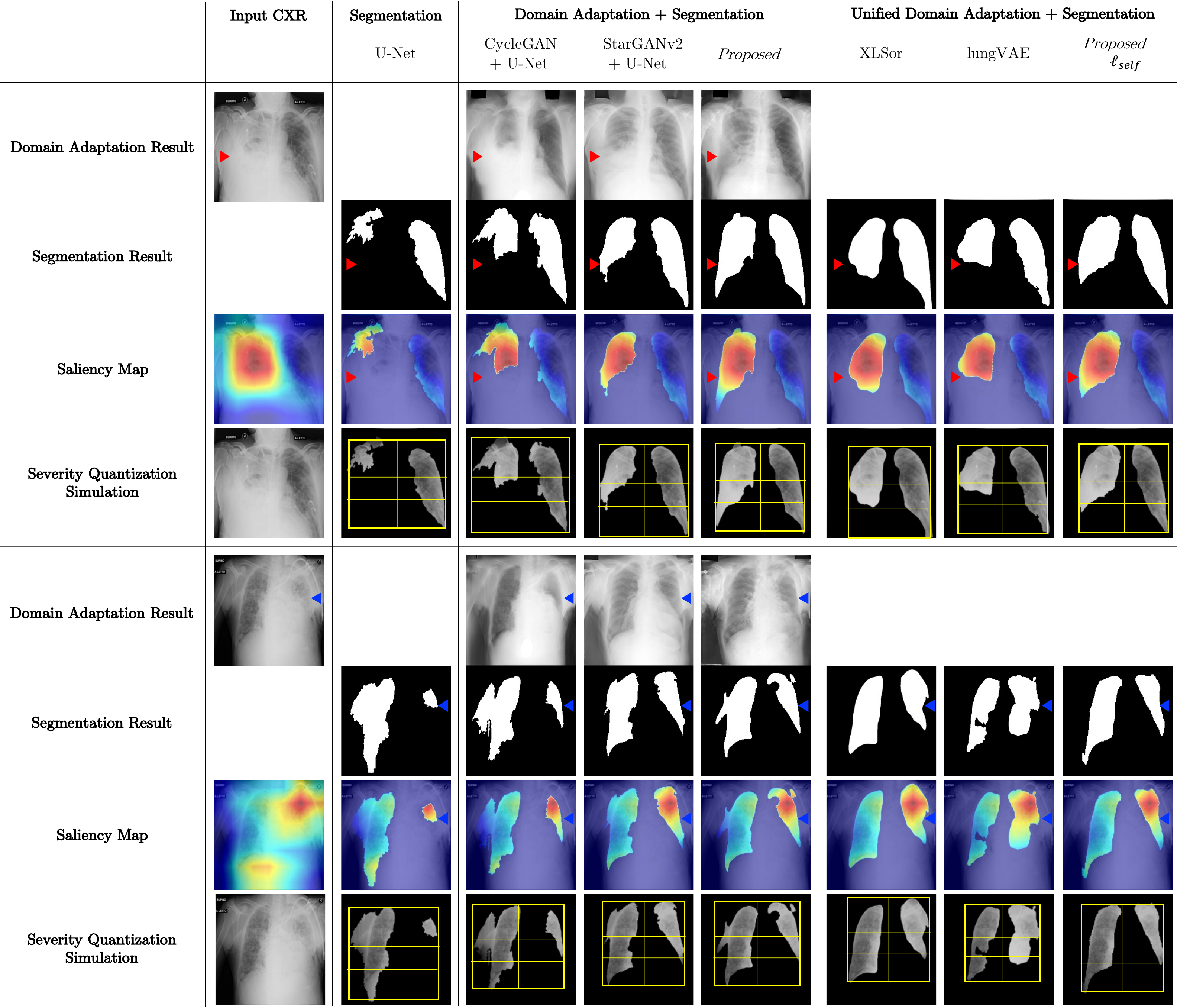}
	\caption{Model performance comparison on various DL-based automatic CXR analysis. Red and blue triangles indicate highly consolidated lung regions in CXRs. Yellow boxes indicate array-based six subdivisions of lung for severity quantification. } 
	\label{fig:component2}
\end{figure}

Fig. \ref{fig:component2} shows that \emph{Proposed} and \emph{Proposed+$\ell_{self}$} models provide the most suitable segmentation mask for saliency map or severity array generation. 
The results demonstrate that our methods can be utilized for various automatic CXR analysis tasks.

\section{Network Architecture}
\label{sp:architecture}

We provide details of the proposed framework, as shown in Table \ref{table:gen}, \ref{table:adain}, \ref{table:style} and \ref{table:dis}. Note that each input and output dimension for the domain adaption task is single channel ($C_{da}$), whereas the segmentation task requires two channels ($C_{seg}$) to separate foreground and background channels.

\begin{table}[hbt!]
	\caption{Generator network architecture.}
	\label{table:gen}
	\centering
	\scalebox{0.8}{
		\begin{tabular}{M{2cm}|M{3cm}M{2cm}M{2cm}M{2.8cm}}
			\hline  
			\multirow{2}{*} {\bf{Module}} &   \multirow{2}{*}{\bf{Layer}} &  \multirow{2}{*}{\bf{Norm}}  &  \multirow{2}{*}{\bf{Resize}} & \bf{Input dimension} \\ 
			& & &  & (C $\times$ H $\times$ W )  \\
			\hline
			\hline
			{In} & Conv 1$\times$1 & - &  - & $C_{da}$ $\times $ 256 $ \times $ 256 \\
			\hline
			\multirow{2}{*}{{Encoder} }& ResBlock $\times$ 4 & AdaIN & Down & 64 $\times$ 256 $\times$ 256 \\
			& ResBlock $\times$ 2 & AdaIN & - &  512 x 16 x 16 \\
			\hline
			\multirow{2}{*}{Decoder } & ResBlock $\times$ 2 & AdaIN & - & 512 $\times$ 16 $\times$ 16 \\
			& ResBlock $\times$ 4 & AdaIN & Up & 512 $\times$ 16 $\times$ 16 \\
			\hline
			\multirow{3}{*}{Unshared}  & Norm & IN & - &  \multirow{3}{*} {64 $\times$ 256 $\times$ 256} \\
			& Leaky ReLU & - & - &  \\
			& Conv 1$\times$1  & - & - &  \\
			\hline 
			\hline
			\multirow{2}{*}{Output}&  \multirow{2}{*}{-} & \multirow{2}{*}{-} &  \multirow{2}{*}{-} & $C_{da}$ $\times $ 256 $ \times $ 256 \\
			&  &  &  & $C_{seg}$ $\times $ 256 $ \times $ 256 \\
			\hline  
	\end{tabular}  }
\end{table}

\begin{table}[hbt!]
	\caption{AdaIN code generator architecture.}
	\label{table:adain}
	\centering
	\scalebox{0.8}{
		\begin{tabular}{M{2cm}|M{2cm}M{2.8cm}}
			\hline  
			\multirow{2}{*} {\bf{Module}} &   \multirow{2}{*}{\bf{Layer}} & \bf{Input dimension} \\ 
			& &  (C)  \\
			\hline
			\hline
			{In} & Latent z & 4 \\
			\hline
			\multirow{2}{*}{{Shared} }& Linear $\times$ 3 & 4 \\
			& Linear  $\times$ 1 & 512 \\
			\hline
			\multirow{2}{*}{Unshared}  & Linear $\times$ 3 & 512 \\
			& Linear  $\times$ 1 &  512 \\
			\hline 
			\hline
			Output & - & 16 $ \times$ K \\
			\hline  
	\end{tabular}  }
\end{table}

\begin{table}[hbt!]
	\caption{Style encoder architecture.}
	\label{table:style}
	\centering
	\scalebox{0.8}{
		\begin{tabular}{M{2cm}|M{2.5cm}M{2.cm}M{2cm}}
			\hline  
			\multirow{2}{*} {\bf{Module}} &   \multirow{2}{*}{\bf{Layer}} & \bf{Input channel}  & \bf{Input size}\\ 
			& & (C)   & (H $\times$ W )   \\
			\hline
			\hline
			{Unshared} & Conv 1$\times$1  &  $C_{da}$, $C_{seg}$  & 256 $ \times $ 256  \\
			\hline
			\multirow{4}{*}{{Shared} }&  ResBlock $\times$ 6 & 64 & {256 $ \times $ 256} \\
			& Leaky ReLU & 512 & {4 $ \times $ 4}  \\
			& Conv 4$\times$4  & 512 & {4 $ \times $ 4}  \\
			& Leaky ReLU & 512 & {1 $ \times $ 1}  \\
			\hline
			{Unshared}  & Linear & 512 & {1 $ \times $ 1}  \\
			\hline 
			\hline
			Output & - & 16  $ \times$ K& -\\
			\hline  
			
	\end{tabular}  }
\end{table}

\begin{table}[hbt!]
	\caption{Discriminator architecture.}
	\label{table:dis}
	\centering
	\scalebox{0.8}{
		\begin{tabular}{M{2cm}|M{2.5cm}M{2.cm}M{2cm}}
			\hline  
			\multirow{2}{*} {\bf{Module}} &   \multirow{2}{*}{\bf{Layer}} & \bf{Input channel}  & \bf{Input size}\\ 
			& & ($C_{da}$)   & (H $\times$ W )   \\
			\hline
			\hline
			\multirow{5}{*}{{Shared} }&  Conv 1$\times$1  &  1 & 256 $ \times $ 256  \\
			&  ResBlock $\times$ 6 & 64 & {256 $ \times $ 256} \\
			& Leaky ReLU & 512 & {4 $ \times $ 4}  \\
			& Conv 4$\times$4  & 512 & {4 $ \times $ 4}  \\
			& Leaky ReLU & 512 & {1 $ \times $ 1}  \\
			\hline
			{Unshared}  & Conv 1$\times$1 & 512 & {1 $ \times $ 1}  \\
			\hline 
			\hline
			Output & - &  1 $ \times$ K & -\\
			\hline  
	\end{tabular}  }
\end{table}

\section{Adaptive Instance Normalization}
\label{sp:adain}

AdaIN has been proposed as an extension of the instance normalization \cite{huang2017arbitrary}.
AdaIN layer receives a content input $x$ and a AdaIN code $a$, and simply aligns the channel-wise mean and variance of $x$ to match those of desirable style by:

\begin{align}
	AdaIN = f(a)\Big(\frac{x-\mu(x)}{\sigma(x)}\Big) + g(a)
	\label{eq:adain}
\end{align}
where $f$ and $g$ compute affine parameters from AdaIN code $a$, $\mu$ and $\sigma$ represent mean and variance, respectively. In this way, AdaIN simply scales the normalized content input with $\sigma(a)$, and shifts with $\mu(a)$.

\section{Ablation Study}
\label{sp:ablation}

{For ablation study, we analyzed contribution of different losses for the supervised segmentation task.} 
We cumulatively added each loss to the baseline model, and compared segmentation performance on abnormal CXR.

%
%

\begin{figure}[hbt!]
\centering
\includegraphics[width=12cm]{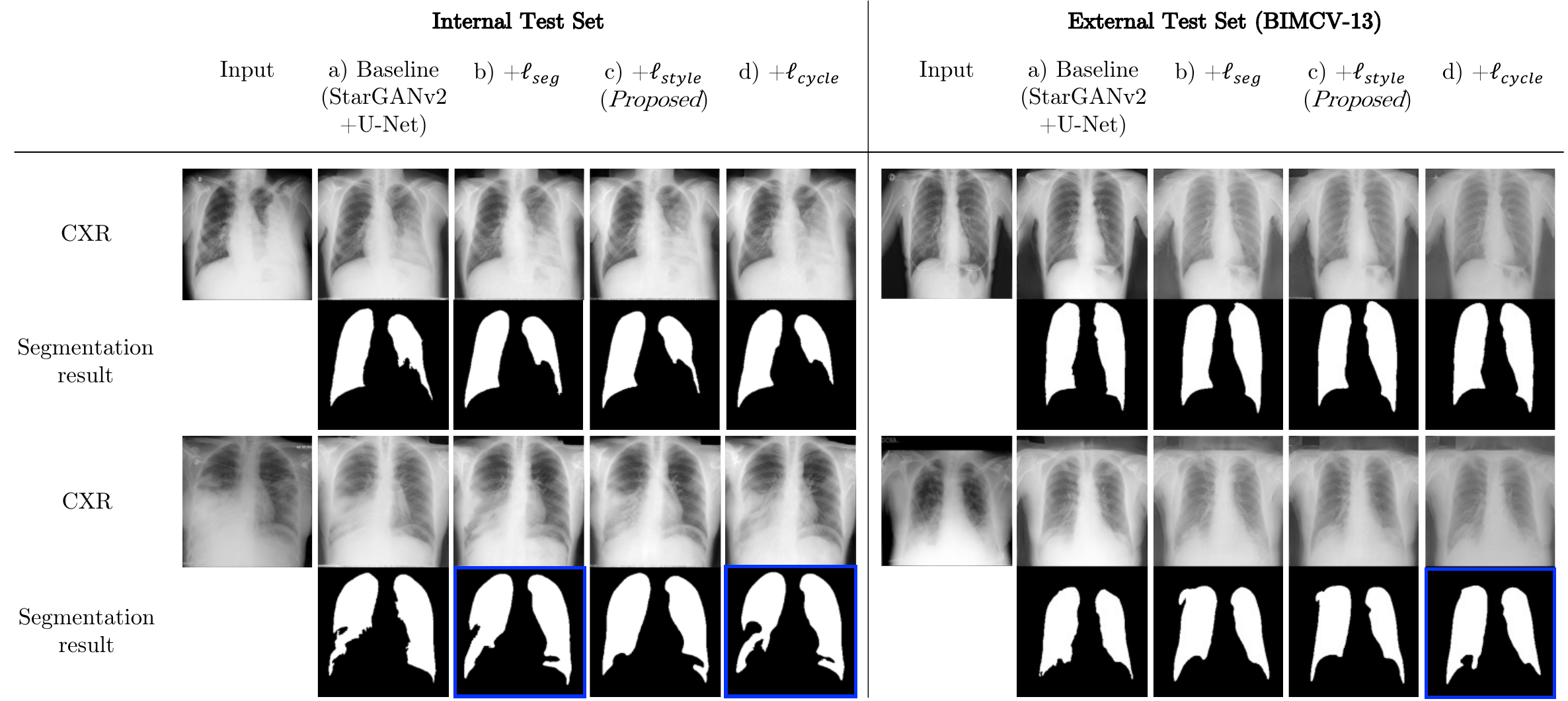}
\caption{Ablation study with different losses for the segmentation task.}
\label{fig:component}
\end{figure}

Fig. \ref{fig:component} compares performance of each configuration. 
In configuration (b) with additional segmentation loss, domain adaptation performance was superior to the baseline. However, we observed that the segmentation results have several concave regions (blue boxes), which failed to resemble the general shape of lung structure. 
In configuration (c) with additional style loss, we observed that the segmentation results resemble normal lung better than (b), thanks to the style loss, which can extract common features of normal and abnormal CXR via the shared layer of the style encoder. 
In configuration (d) with additional cycle-consistency loss, we observed rather degraded lung segmentation performance as depicted as blue boxes, which have more concave regions compared to that of (c). 
The cycle-consistency loss, which tries to revert the generated lung mask back to the original image, may disturb the network to segment extremely consolidated lung regions. 

Based on the ablation study results, we set the configuration (c) as our supervised segmentation loss.

\section{Domain Adaptation Loss}
\label{sp:da}

{Domain adaptation loss basically follows StarGANv2 \cite{choi2020stargan}, given by}
\begin{align}%
&\ell_{da}(G,F_e,F_d, S, D)  = \label{eq:da_supple} \\
&~~\ell_{adv}(G,D,F_d, S)   \\
&+\lambda_{cycle} \ell_{cycle}(G, S)  \\
&+\lambda_{style} \ell_{style}(G, S)   \\
&-\lambda_{div} \ell_{div}(G, F_d, S), 
\end{align}
where $\lambda_{cycle}$,$\lambda_{style}$  and $\lambda_{div}$ are hyper-parameters and $\ell_{adv}$ is the adversarial loss defined by
\begin{align}
&\ell_{adv}(G,D,F_d, S) = \mathbb{E}_{s\sim P_{\Sc}}
\left[\log{D_{\Sc}}(s) \right] 
+  \mathbb{E}_{s\sim P_{\Sc}}
\left[\log(1 - {D_{\Tc}}(G(s, a_{da}^{\Tc})) \right],
\label{eq:adv2}
\end{align}
{where $\Sc$ and $\Tc$ are source and target domains, which are chosen randomly from $\Xc$ and $\Yc$ so that all domain combinations can be considered.
Furthermore, the learnable part of the AdaIN code $a_{da}^{\Tc}$ is generated either from the encoder AdaIN coder generator $F_d$ or the style encoder $S$ given a reference target $t\in \Tc$.

The cycle-consistency loss $\ell_{cycle}$ is defined as follows:
\begin{align}
\ell_{cycle}(G, S) 
= \mathbb{E}_{s\sim P_\Sc} 
\left[ \lVert x  - G(G(s, a_{da}^{\Tc}), a_{da}^{\Sc})
\rVert_{1} \right]  \  .
\label{eq:cycle}
\end{align}

Similar to the cycle-consistency loss $\ell_{cycle}$  for the images, we introduce the style loss $\ell_{style}$  to enforce the cycle-consistency in the AdaIN code domain.
More specifically, once a fake image is generated using a domain-specific AdaIN code, the style encoder with the fake image as an input should reproduce the original AdaIN code.
This can be achieved by minimizing the following style loss:
\begin{align}
&\ell_{style}(G, S) 
= \mathbb{E}_{s\sim P_{\Sc}} 
\left[ \lVert a_{da}^{\Tc} - {S}(G(s, a_{da}^{\Tc}))
\rVert_{1} \right].
\end{align}

Finally, to make the generated fake images  diverse, the difference between two fake images that are generated by two different AdaIN codes should be maximized. This can be achieved by maximizing the following loss: 
\begin{align}
&\ell_{div} (G, F_d, S)
= \mathbb{E}_{s\sim P_{\Sc}} 
\left[ \lVert G(s, a_{da}^{\Tc})
- G(s, {a'}_{da}^{\Tc})
\rVert_{1} \right],
\label{eq:diversity}
\end{align}
where an additional ${a'}_{da}^{\Tc}$ is generated either from the encoder AdaIN coder generator $F_d$ or the style encoder $S$ given an additional reference image.

\section{Implementation Details}
\label{sp:mplementation}

The proposed method was implemented with PyTorch library \cite{paszke2017automatic}. We applied Adam optimizer \cite{kingma2014adam} to train the models and set the batch size 1. The model was trained using a NVIDIA GeForce GTX 1080 Ti GPU.
Hyper parameters were chosen to be $\lambda_{cycle}=2$, $\lambda_{style}=1$, $\lambda_{div}=1$, $\lambda_{seg}=5$, $\lambda_{inter}=10$ and $\lambda_{intra}=1$. Learning rate was optimized to $0.0001$. Once training iteration reaches {certain fixed iteration points} throughout the total iterations, the learning rate was reduced by factor of 10.

The network was trained for 20K iterations to simultaneously train domain adaptation and supervised segmentation tasks. We adopted early stopping strategy based on validation performance of abnormal chest X-ray radiograph (CXR) segmentation results. In terms of the training sequence, the self-supervised training started after training the domain adaptation and supervised segmentation tasks until they guaranteed certain performances. 
For self-supervised learning, the network was continued to be trained in self-supervised manner for additional 5K iterations. 

{At the inference phase, as for post-processing steps, two largest contours were automatically selected based on contour areas, and any holes within each contour were filled. The post-processing technique was identically applied to all the comparative model outputs for fair comparison.}

\section{Comparative Model Implementations}
\label{sp:compare}

{
For comparative study, baseline models for domain adaptation and supervised segmentation tasks, i.e., CycleGAN \cite{Zhu_2017_ICCV}, MUNIT \cite{huang2018multimodal}, StarGANv2 \cite{choi2020stargan} and U-Net \cite{ronneberger2015unet}, were trained with identical conditions to that of the proposed model. 
For comparing performance of the unified domain adaptation and segmentation network, we inferenced pre-trained networks optimized for abnormal CXR segmentation, i.e., XLSor \cite{tang2019XLSor} and lungVAE \cite{selvan2020lung}, by utilizing their official source codes. \footnote{\url{https://github.com/raghavian/lungVAE}}\footnote{\url{https://github.com/rsummers11/CADLab/tree/master/Lung_Segmentation_XLSor}}
}


\section{Error Analysis}
\label{sp:error}
We analyzed typical error cases, which failed to be segmented, and the error cases were grouped into three categories: (a) Over-segmentation on background pixels, (b) distorted lung shape, and (c) distorged lung boundary, as shown in Fig. \ref{fig:error}. 

\begin{figure}[hbt!]
\centering
\includegraphics[width=12cm]{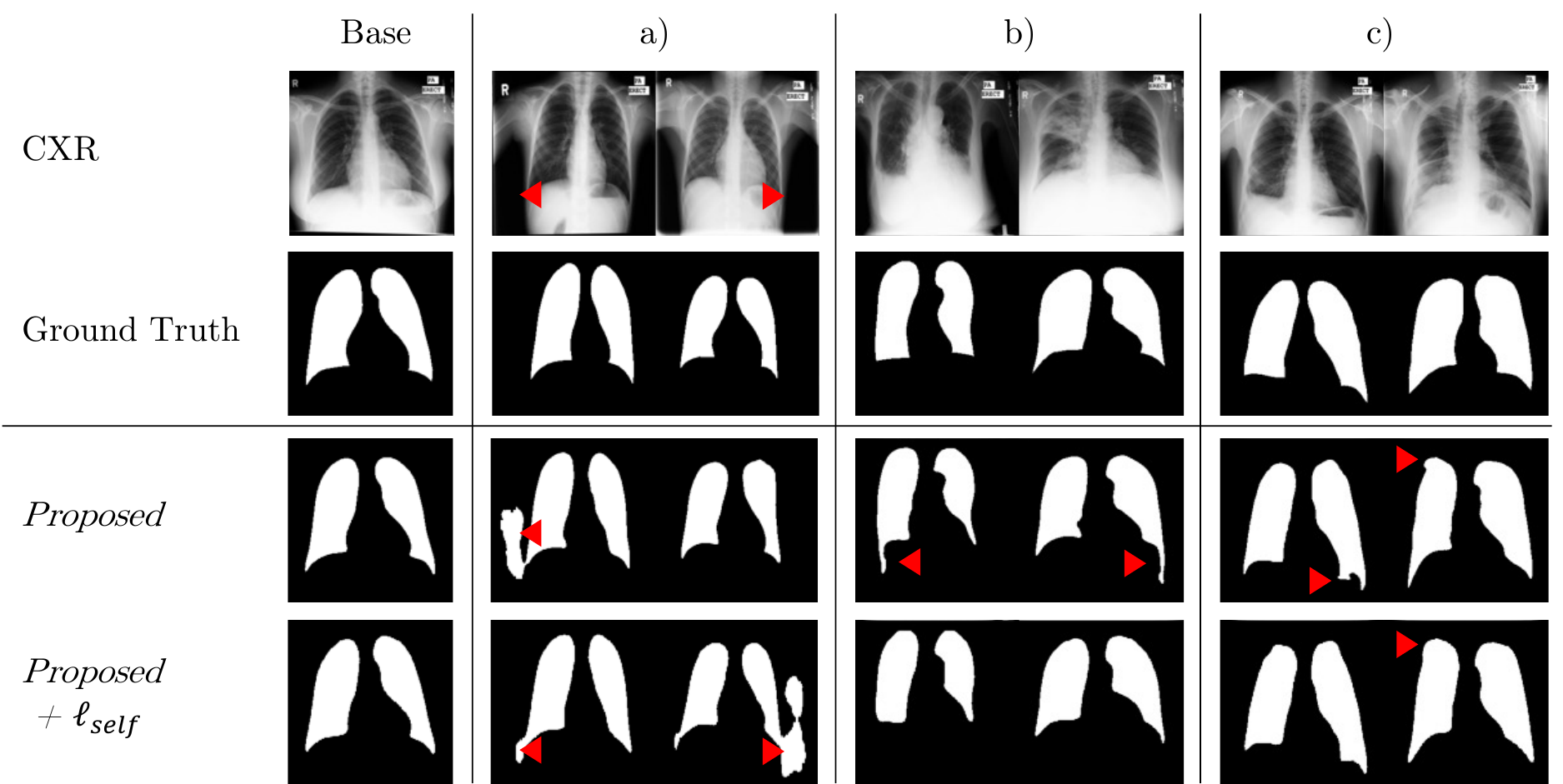}
\caption{Representative error cases. (a) Over-segmentation on background pixels, (b) distorted lung shape and (c) distorted lung boundary. Base indicates standard segmentation result.}
\label{fig:error}
\end{figure}

\section{Computational Costs}
\label{sp:costs}

The proposed unified framework costs less training computation resources, compared to training individual domain adaptation and segmentation networks. Table \ref{table:cost} shows total network parameters utilized for either training or inference of comparative networks. 

Once the model is trained, only the generator with pre-built AdaIN codes are used at the inference phase, thus the model costs only the single generator. 

{As shown in Table \ref{table:cost}, \emph{Proposed} and \emph{Proposed+$\ell_{self}$} models need the least number of network parameters, with most promising segmentation performance.}
{Specifically, compared to \emph{Proposed} model, \emph{Proposed+$\ell_{self}$} model only needs a single inference without preceding domain adaptation task, with comparable segmentation performance to that of \emph{Proposed} model.}

\begin{table}[h!]
\caption{Number of trainable and inference parameters.}
\label{table:cost}
\centering
\scalebox{0.8}{
	\begin{tabular}{l|ccc|ccc|c}
		\hline  
		\multirow{2}{*}{\textbf{Method}} 
		~& \multicolumn{3}{c|}{\textbf{Training}} & \multicolumn{3}{c|}{\textbf{Inference(S)}} & {\textbf{Inference}} \\ 
		& {Generator(S)} & {          Others          } & {          Total          } & {Generator(S)} & {          Others          } & {          Total          } & {Time}  \\
		\hline
		\hline
		\textbf{SS} & & & & &  &  & \\
		U-Net \cite{ronneberger2015unet} & 29M &  - & 29M & 29M &  - & 29M & $\times$ 1 \\
		XLSor \cite{tang2019XLSor}  & 71M & -  & 71M & 71M & -  & 71M  & $\times$ 1 \\
		\hline
		\textbf{DA} & & & & &  &  &  \\
		CycleGAN \cite{Zhu_2017_ICCV} & -& 29M  & 29M & - & -  & -  \\
		MUNIT \cite{huang2018multimodal}  & - & 47M  & 47M  & - & -  & -  \\
		StarGANv2 \cite{choi2020stargan}  & - &  78M & 78M  & - & -  & -   \\
		\hline
		\textbf{DA+SS} & & & & &  &  & \\
		CycleGAN + U-Net & 29M & 29M & 58M & 29M & 11M & 40M & $\times$ 2 \\
		{StarGANv2 + U-Net} & 29M &  78M & 127M &  29M &  34M & 63M  & $\times$ 2 \\
		{\emph{Proposed } } & 34M &  45M & 79M & 34M &  - & \textbf{34M} & $\times$ 2\\
		\hline
		\textbf{UDS/Self} & & & &  &  & & \\
		MUNIT + XLSor & 71M & 47M & 118M & 71M & - & 71M & $\times$ 1 \\ 
		lungVAE \cite{selvan2020lung}	 & 34M & - & 34M & 34M & - & \textbf{34M} & $\times$ 1\\
		{\emph{Proposed+$\ell_{self}$}} & 34M &  46M & 80M & 34M & - & \textbf{34M} & $\times$ 1 \\
		\hline  
		\multicolumn{8}{l}{\textit{Note:} SS, supervised segmentation; DA, domain adaptation; UDS, unified DA+SS; Self, self-}\\
		\multicolumn{8}{l}{supervised segmentation; (S), segmentation task; Others, other module parameters.}\\
\end{tabular}  }
\end{table}

\end{document}